\documentclass[%
 reprint,
superscriptaddress,
showpacs,
 amsmath,amssymb,
 aps,
]{revtex4-1}

\usepackage{graphicx}% Include figure files
\usepackage{dcolumn}% Align table columns on decimal point
\usepackage{bm}% bold math
\usepackage{amsmath}
\usepackage{autobreak}
\usepackage{hyperref}% add hypertext capabilities
\usepackage[mathlines]{lineno}% Enable numbering of text and display math
\usepackage{natbib}
\usepackage{multirow}
\usepackage{booktabs}

\begin{document}

\preprint{APS/123-QED}

\title{Constraints on sub-GeV dark matter boosted by cosmic rays from the \\CDEX-10 experiment at the China Jinping Underground Laboratory}

\affiliation{Key Laboratory of Particle and Radiation Imaging (Ministry of Education) and Department of Engineering Physics, Tsinghua University, Beijing 100084}
\affiliation{Institute of Physics, Academia Sinica, Taipei 11529}

\affiliation{Department of Physics, Tsinghua University, Beijing 100084}
\affiliation{NUCTECH Company, Beijing 100084}
\affiliation{YaLong River Hydropower Development Company, Chengdu 610051}
\affiliation{College of Nuclear Science and Technology, Beijing Normal University, Beijing 100875}
\affiliation{College of Physics, Sichuan University, Chengdu 610065}
\affiliation{School of Physics, Peking University, Beijing 100871}
\affiliation{Department of Nuclear Physics, China Institute of Atomic Energy, Beijing 102413}
\affiliation{Sino-French Institute of Nuclear and Technology, Sun Yat-sen University, Zhuhai 519082}
\affiliation{School of Physics, Nankai University, Tianjin 300071}
\affiliation{Department of Physics, Banaras Hindu University, Varanasi 221005}
\affiliation{Department of Physics, Beijing Normal University, Beijing 100875}

\author{R.~Xu}
\affiliation{Key Laboratory of Particle and Radiation Imaging (Ministry of Education) and Department of Engineering Physics, Tsinghua University, Beijing 100084}

\author{L.~T.~Yang}\altaffiliation [Corresponding author: ]{yanglt@mail.tsinghua.edu.cn}
\affiliation{Key Laboratory of Particle and Radiation Imaging (Ministry of Education) and Department of Engineering Physics, Tsinghua University, Beijing 100084}
\author{Q. Yue}\altaffiliation [Corresponding author: ]{yueq@mail.tsinghua.edu.cn}
\affiliation{Key Laboratory of Particle and Radiation Imaging (Ministry of Education) and Department of Engineering Physics, Tsinghua University, Beijing 100084}

\author{K.~J.~Kang}
\affiliation{Key Laboratory of Particle and Radiation Imaging (Ministry of Education) and Department of Engineering Physics, Tsinghua University, Beijing 100084}
\author{Y.~J.~Li}
\affiliation{Key Laboratory of Particle and Radiation Imaging (Ministry of Education) and Department of Engineering Physics, Tsinghua University, Beijing 100084}
\author{M. Agartioglu}
\altaffiliation{Participating as a member of TEXONO Collaboration}
\affiliation{Institute of Physics, Academia Sinica, Taipei 11529}
\author{H.~P.~An}
\affiliation{Key Laboratory of Particle and Radiation Imaging (Ministry of Education) and Department of Engineering Physics, Tsinghua University, Beijing 100084}
\affiliation{Department of Physics, Tsinghua University, Beijing 100084}
\author{J.~P.~Chang}
\affiliation{NUCTECH Company, Beijing 100084}

\author{Y.~H.~Chen}
\affiliation{YaLong River Hydropower Development Company, Chengdu 610051}
\author{J.~P.~Cheng}
\affiliation{Key Laboratory of Particle and Radiation Imaging (Ministry of Education) and Department of Engineering Physics, Tsinghua University, Beijing 100084}
\affiliation{College of Nuclear Science and Technology, Beijing Normal University, Beijing 100875}
\author{W.~H.~Dai}
\affiliation{Key Laboratory of Particle and Radiation Imaging (Ministry of Education) and Department of Engineering Physics, Tsinghua University, Beijing 100084}
\author{Z.~Deng}
\affiliation{Key Laboratory of Particle and Radiation Imaging (Ministry of Education) and Department of Engineering Physics, Tsinghua University, Beijing 100084}

\author{C.~H.~Fang}
\affiliation{College of Physics, Sichuan University, Chengdu 610065}

\author{X.~P.~Geng}
\affiliation{Key Laboratory of Particle and Radiation Imaging (Ministry of Education) and Department of Engineering Physics, Tsinghua University, Beijing 100084}
\author{H.~Gong}
\affiliation{Key Laboratory of Particle and Radiation Imaging (Ministry of Education) and Department of Engineering Physics, Tsinghua University, Beijing 100084}
\author{X.~Y.~Guo}
\affiliation{YaLong River Hydropower Development Company, Chengdu 610051}
\author{Q.~J.~Guo}
\affiliation{School of Physics, Peking University, Beijing 100871}
\author{L. He}
\affiliation{NUCTECH Company, Beijing 100084}
\author{S.~M.~He}
\affiliation{YaLong River Hydropower Development Company, Chengdu 610051}
\author{J.~W.~Hu}
\affiliation{Key Laboratory of Particle and Radiation Imaging (Ministry of Education) and Department of Engineering Physics, Tsinghua University, Beijing 100084}

\author{H.~X.~Huang}
\affiliation{Department of Nuclear Physics, China Institute of Atomic Energy, Beijing 102413}
\author{T.~C.~Huang}
\affiliation{Sino-French Institute of Nuclear and Technology, Sun Yat-sen University, Zhuhai 519082}

\author{H.~T.~Jia}
\affiliation{College of Physics, Sichuan University, Chengdu 610065}
\author{X.~Jiang}
\affiliation{College of Physics, Sichuan University, Chengdu 610065}

\author{H.~B.~Li}
\altaffiliation{Participating as a member of TEXONO Collaboration}
\affiliation{Institute of Physics, Academia Sinica, Taipei 11529}

\author{J.~M.~Li}
\affiliation{Key Laboratory of Particle and Radiation Imaging (Ministry of Education) and Department of Engineering Physics, Tsinghua University, Beijing 100084}
\author{J.~Li}
\affiliation{Key Laboratory of Particle and Radiation Imaging (Ministry of Education) and Department of Engineering Physics, Tsinghua University, Beijing 100084}

\author{Q.~Y.~Li}
\affiliation{College of Physics, Sichuan University, Chengdu 610065}
\author{R.~M.~J.~Li}
\affiliation{College of Physics, Sichuan University, Chengdu 610065}
\author{X.~Q.~Li}
\affiliation{School of Physics, Nankai University, Tianjin 300071}
\author{Y.~L.~Li}
\affiliation{Key Laboratory of Particle and Radiation Imaging (Ministry of Education) and Department of Engineering Physics, Tsinghua University, Beijing 100084}
\author{Y.~F.~Liang}
\affiliation{Key Laboratory of Particle and Radiation Imaging (Ministry of Education) and Department of Engineering Physics, Tsinghua University, Beijing 100084}
\author {B. Liao}
\affiliation{College of Nuclear Science and Technology, Beijing Normal University, Beijing 100875}
\author{F.~K.~Lin}
\altaffiliation{Participating as a member of TEXONO Collaboration}
\affiliation{Institute of Physics, Academia Sinica, Taipei 11529}
\author{S.~T.~Lin}
\affiliation{College of Physics, Sichuan University, Chengdu 610065}
\author{S.~K.~Liu}
\affiliation{College of Physics, Sichuan University, Chengdu 610065}
\author{Y.~Liu}
\affiliation{College of Physics, Sichuan University, Chengdu 610065}
\author {Y.~D.~Liu}
\affiliation{College of Nuclear Science and Technology, Beijing Normal University, Beijing 100875}
\author {Y.~Y.~Liu}
\affiliation{College of Nuclear Science and Technology, Beijing Normal University, Beijing 100875}
\author{Z.~Z.~Liu}
\affiliation{Key Laboratory of Particle and Radiation Imaging (Ministry of Education) and Department of Engineering Physics, Tsinghua University, Beijing 100084}
\author{H.~Ma}
\affiliation{Key Laboratory of Particle and Radiation Imaging (Ministry of Education) and Department of Engineering Physics, Tsinghua University, Beijing 100084}

\author{Y.~C.~Mao}
\affiliation{School of Physics, Peking University, Beijing 100871}
\author{Q.~Y.~Nie}
\affiliation{Key Laboratory of Particle and Radiation Imaging (Ministry of Education) and Department of Engineering Physics, Tsinghua University, Beijing 100084}
\author{J.~H.~Ning}
\affiliation{YaLong River Hydropower Development Company, Chengdu 610051}
\author{H.~Pan}
\affiliation{NUCTECH Company, Beijing 100084}
\author{N.~C.~Qi}
\affiliation{YaLong River Hydropower Development Company, Chengdu 610051}
\author{J.~Ren}
\affiliation{Department of Nuclear Physics, China Institute of Atomic Energy, Beijing 102413}
\author{X.~C.~Ruan}
\affiliation{Department of Nuclear Physics, China Institute of Atomic Energy, Beijing 102413}
\author{K.~Saraswat}
\altaffiliation{Participating as a member of TEXONO Collaboration}
\affiliation{Institute of Physics, Academia Sinica, Taipei 11529}
\author{V.~Sharma}
\altaffiliation{Participating as a member of TEXONO Collaboration}
\affiliation{Institute of Physics, Academia Sinica, Taipei 11529}
\affiliation{Department of Physics, Banaras Hindu University, Varanasi 221005}
\author{Z.~She}
\affiliation{Key Laboratory of Particle and Radiation Imaging (Ministry of Education) and Department of Engineering Physics, Tsinghua University, Beijing 100084}

\author{M.~K.~Singh}
\altaffiliation{Participating as a member of TEXONO Collaboration}
\affiliation{Institute of Physics, Academia Sinica, Taipei 11529}
\affiliation{Department of Physics, Banaras Hindu University, Varanasi 221005}

\author {T.~X.~Sun}
\affiliation{College of Nuclear Science and Technology, Beijing Normal University, Beijing 100875}

\author{C.~J.~Tang}
\affiliation{College of Physics, Sichuan University, Chengdu 610065}
\author{W.~Y.~Tang}
\affiliation{Key Laboratory of Particle and Radiation Imaging (Ministry of Education) and Department of Engineering Physics, Tsinghua University, Beijing 100084}
\author{Y.~Tian}
\affiliation{Key Laboratory of Particle and Radiation Imaging (Ministry of Education) and Department of Engineering Physics, Tsinghua University, Beijing 100084}

\author {G.~F.~Wang}
\affiliation{College of Nuclear Science and Technology, Beijing Normal University, Beijing 100875}

\author{L.~Wang}
\affiliation{Department of Physics, Beijing Normal University, Beijing 100875}
\author{Q.~Wang}
\affiliation{Key Laboratory of Particle and Radiation Imaging (Ministry of Education) and Department of Engineering Physics, Tsinghua University, Beijing 100084}
\affiliation{Department of Physics, Tsinghua University, Beijing 100084}
\author{Y.~Wang}
\affiliation{Key Laboratory of Particle and Radiation Imaging (Ministry of Education) and Department of Engineering Physics, Tsinghua University, Beijing 100084}
\affiliation{Department of Physics, Tsinghua University, Beijing 100084}
\author{Y.~X.~Wang}
\affiliation{School of Physics, Peking University, Beijing 100871}

\author{H.~T.~Wong}
\altaffiliation{Participating as a member of TEXONO Collaboration}
\affiliation{Institute of Physics, Academia Sinica, Taipei 11529}
\author{S.~Y.~Wu}
\affiliation{YaLong River Hydropower Development Company, Chengdu 610051}
\author{Y.~C.~Wu}
\affiliation{Key Laboratory of Particle and Radiation Imaging (Ministry of Education) and Department of Engineering Physics, Tsinghua University, Beijing 100084}
\author{H.~Y.~Xing}
\affiliation{College of Physics, Sichuan University, Chengdu 610065}
\author{Y.~Xu}
\affiliation{School of Physics, Nankai University, Tianjin 300071}
\author{T.~Xue}
\affiliation{Key Laboratory of Particle and Radiation Imaging (Ministry of Education) and Department of Engineering Physics, Tsinghua University, Beijing 100084}

\author{Y.~L.~Yan}
\affiliation{College of Physics, Sichuan University, Chengdu 610065}
\author{C.~H.~Yeh}
\altaffiliation{Participating as a member of TEXONO Collaboration}
\affiliation{Institute of Physics, Academia Sinica, Taipei 11529}
\author{N.~Yi}
\affiliation{Key Laboratory of Particle and Radiation Imaging (Ministry of Education) and Department of Engineering Physics, Tsinghua University, Beijing 100084}
\author{C.~X.~Yu}
\affiliation{School of Physics, Nankai University, Tianjin 300071}
\author{H.~J.~Yu}
\affiliation{NUCTECH Company, Beijing 100084}
\author{J.~F.~Yue}
\affiliation{YaLong River Hydropower Development Company, Chengdu 610051}
\author{M.~Zeng}
\affiliation{Key Laboratory of Particle and Radiation Imaging (Ministry of Education) and Department of Engineering Physics, Tsinghua University, Beijing 100084}
\author{Z.~Zeng}
\affiliation{Key Laboratory of Particle and Radiation Imaging (Ministry of Education) and Department of Engineering Physics, Tsinghua University, Beijing 100084}
\author{B.~T.~Zhang}
\affiliation{Key Laboratory of Particle and Radiation Imaging (Ministry of Education) and Department of Engineering Physics, Tsinghua University, Beijing 100084}
\author {F.~S.~Zhang}
\affiliation{College of Nuclear Science and Technology, Beijing Normal University, Beijing 100875}
\author{L. Zhang}
\affiliation{College of Physics, Sichuan University, Chengdu 610065}
\author{Z.~H.~Zhang}
\affiliation{Key Laboratory of Particle and Radiation Imaging (Ministry of Education) and Department of Engineering Physics, Tsinghua University, Beijing 100084}
\author{Z.~Y.~Zhang}
\affiliation{Key Laboratory of Particle and Radiation Imaging (Ministry of Education) and Department of Engineering Physics, Tsinghua University, Beijing 100084}
\author{K.~K.~Zhao}
\affiliation{College of Physics, Sichuan University, Chengdu 610065}
\author{M.~G.~Zhao}
\affiliation{School of Physics, Nankai University, Tianjin 300071}
\author{J.~F.~Zhou}
\affiliation{YaLong River Hydropower Development Company, Chengdu 610051}
\author{Z.~Y.~Zhou}
\affiliation{Department of Nuclear Physics, China Institute of Atomic Energy, Beijing 102413}
\author{J.~J.~Zhu}
\affiliation{College of Physics, Sichuan University, Chengdu 610065}

\collaboration{CDEX Collaboration}
\noaffiliation

\date{\today}

\begin{abstract}
We present new constraints on light dark matter boosted by cosmic rays (CRDM) using the 205.4 kg day data of the CDEX-10 experiment conducted at the China Jinping Underground Laboratory. The Monte Carlo simulation package $\tt CJPL\_ESS$ was employed to evaluate the Earth shielding effect. Several key factors have been introduced and discussed in our CRDM analysis, including the contributions from heavier CR nuclei than proton and helium, the inhomogeneity of CR distribution, and the impact of the form factor in the Earth attenuation calculation. Our result excludes the dark matter--nucleon elastic scattering cross section region from $1.7\times 10^{-30}$ to $10^{-26}~\rm cm^2$ for dark matter of 10 keV$/c^2$ to 1 GeV$/c^2$.

\end{abstract}
\maketitle

\section{\label{sec1}Introduction}
Compelling cosmological evidence indicates the existence of dark matter (DM, denoted as $\chi$) in the
Universe~\cite{pdg2018,bertone_particle_2005}. Myriad efforts have been pursued, including collider searches, direct detection (DD) experiments, and a wide range of astrophysical and cosmological studies, yet no clear signals have been observed to date. 

Collider searches can be used to produce exclusion regions that provide upper bounds on the cross section that weaken at lower DM masses~\cite{daci_simplified_2015}. The current constraints in DM detection channels and experimental sensitivities indicate that a large parameter space remains to be explored. DD experiments, such as XENON~\cite{xenon1t}, LUX~\cite{lux}, PandaX~\cite{PandaX}, DarkSide~\cite{darkside}, CRESST~\cite{cresst}, SuperCDMS~\cite{cdmslite}, CoGeNT~\cite{cogent2013}, and CDEX~\cite{cdex1,cdex0,cdex12014,cdex12016,cdex1b2018,cdex102018,cdex10_tech,cdex1b_am,cdex10_eft}, 
are based on DM-nucleus ($\chi$-N) elastic scattering through spin-independent (SI) and spin-dependent interactions. DD experiments rapidly lose sensitivity toward the sub-GeV range, because light DM particles carry insufficient energy to produce nuclear recoil signals with higher energies than the detector threshold. For a DM particle with mass $m_{\chi} \sim 1 $ GeV and a traditional DM escape velocity of 540 km/s, the maximal recoil energy translated to a target Ge nucleus via an elastic scattering interaction is of $\mathcal{O}(10) $ eV, which is considerably lower than the typical detection threshold of a detector searching for DM directly. Cosmological studies have set limits in the sub-GeV range. For example, constraints from the cosmic microwave background (CMB) are approximately $\mathcal{O}(10^{-27})~\rm cm^2$~\cite{cmb} and those from the structure formation are approximately $\mathcal{O}(10^{-29})~\rm cm^2$ in MeV to GeV mass range~\cite{cosmo11}. Stringent constraints can arise from the big bang nucleosynthesis (BBN) if DM particles are relativistic and reach thermal equilibrium in the BBN era. DM with mass below a few MeV can be excluded for some benchmark models~\cite{bbn}.

In the standard halo model (SHM), the DM particles follow a Maxwell-Boltzmann distribution with the most probable velocity of $220$ km/s and a cutoff at the Galactic escape velocity of $540$ km/s~\cite{SHM,shm2}. Recently, it has been realized that light DM particles can be boosted to relativistic or near relativistic momenta via the elastic scattering with cosmic rays (CRs) in the Milky Way halo
~\cite{CRDM_prl,CRDM_TJ,CRDM_ZYF,CRDM_PRD,CRDM_neutrino,CRDM_prd2,PandaX-CRDM}. These secondary DM component particles (referred to as CRDM) will be scattered off in the detectors and transfer sufficient energies to the target nuclei, generating signals that surpass the detection threshold. Considering the CRDM, the lower reach of $m_{\chi}$ in DD experiments can be substantially extended to $\mathcal{O}(100)$ keV range~\cite{CRDM_prl,CRDM_TJ,CRDM_ZYF,CRDM_neutrino,CRDM_PRD,CRDM_prd2,PandaX-CRDM}.

CDEX-10~\cite{cdex102018,cdex10_tech}, the second phase of the CDEX experiment aiming at light DM searches, runs a 10 kg $p$-type point contact germanium (PPCGe)~\cite{soma2016} detector array in the China Jinping Underground Laboratory (CJPL) with about 2400 m of rock overburden~\cite{cjpl}. The detector array consists of three triple-element PPCGe detector strings (C10A, B, and C), which are directly immersed in liquid nitrogen (L$\rm N_2$) for cooling and shielding~\cite{cdex102018}. The 20 cm thick high-purity oxygen-free copper in the L$\rm N_2$ cryostat serves as a passive shield against ambient radioactivity. The L$\rm N_2$ cryostat operates in the polyethylene room with 1 m thick walls at CJPL-I. The configuration of the detector system was described in detail previously~\cite{cdex102018,cdex10_tech}. Since February 2017, the detector has been under stable data-taking conditions. Limits on $\chi$-N SI scattering down to $m_{\chi}\sim2\ \rm GeV/c^2$ were derived at an energy threshold of 160 eVee (electron equivalent energy) with an exposure of 102.8 kg day~\cite{cdex102018}. Recent constraints on the dark photon effective mixing parameter were derived with an exposure of 205.4 kg day~\cite{cdex_darkphoton}.

In this paper, within the CRDM scenario, we reanalyzed the 205.4 kg day dataset from CDEX-10 experiment~\cite{cdex_darkphoton} to set constraints on the $\chi$-nucleon SI interactions, the cross section of which is denoted by $\sigma_{\chi N}$. The Earth shielding effect~\cite{earthshielding,earthshielding1,earthshielding2,earthshielding3,earthshielding4,cjpless}, considering the influence of the rock overburden during the dark matter transportation to an underground laboratory, is utilized to set the upper bound of the exclusion region. We discuss the CR inputs and parameter choices in CRDM studies and examine the related uncertainties. 

\section{\label{sec2}CRDM flux in space} 
In this section, we introduce the CR data used in this work and calculate the CRDM flux in space. Furthermore, a discussion on the effective distance $D_{eff}$ is provided.

\subsection{\label{sec_CR}Cosmic rays in the Galaxy}
Cosmic rays are high-energy particles including protons, electrons, and a range of heavier nuclei, which may originate from and be accelerated by supernova remnants~\cite{cr_origin,cr_origin2}. The Galactic magnetic fields trap the electrically charged CR components in the Milky Way halo and make the CR distribution highly isotropic. Current space-based experiments, such as AMS~\cite{ams1,ams2},
CREAM~\cite{CREAM}, DAMPE~\cite{DAMPE}, CALET~\cite{CALET}, and PAMELA~\cite{pamela}, provide directly measured CR flux data in the MeV--TeV range. The observed CR spectra near Earth peak at $\sim$1 GeV due to solar modulation. Voyager 1 has detected the local interstellar spectra of various CR components away from the heliosphere down to a few MeV, providing results many magnitudes greater than those measured at Earth~\cite{voyager}. The CR fluxes of proton and helium near Earth, which is affected by solar modulation, can be parametrized and described using a fitting formula. The local differential intensity $dI/dR$ of proton and helium as a function of rigidity $R=p/q$ can be expressed by
\begin{equation} \label{fitting}
   \frac{dI}{dR}\times R^{2.7}=\left\{
   \begin{array}{lc}
    \sum\limits_{i=0}^{5} a_i R^i,~\qquad\quad 0.2 \ \rm GV \le R \leq 1\ \rm GV& \\
    b+\dfrac{c}{R}+\dfrac{d_1}{d_2+R}+\dfrac{e_1}{e_2+R} \\+\dfrac{f_1}{f_2+R}+gR,~\qquad\qquad R>1 \ \rm GV&
   \end{array}
   \right.,
\end{equation}
where $a_i,b,c,d_i,e_i,f_i$ and $g$ are the numerical coefficients summarized in Table~\ref{tab:table1}~\cite{boschini_2017,Boschini_2018}. The differential flux as a function of kinetic energy $T$ can be obtained by $\frac{d\Phi}{dT}=4\pi\frac{dI}{dR}\frac{dR}{dT}$. The relativistic $p-T$ relation should be used to derive $dR/dT$. We note for comparison that, in previous CRDM studies such as Ref.~\cite{CRDM_prl}, the fluxes of proton and helium were calculated using Eq.~\ref{fitting}. Extrapolation of the fitting formula below $0.2 \ \rm GV$ for proton and a cutoff with rigidity below $0.2 \ \rm GV$ for helium were made.

\begin{table*}[!htbp]
\caption{\label{tab:table1}
Numerical coefficients in the fitting formula of the differential intensities of proton and helium. Data were obtained from Refs.~\cite{boschini_2017,Boschini_2018}.} 
\begin{ruledtabular}
\begin{tabular}{lccccccccccccccc}
          &  \multicolumn{1}{c}{$a_0$}     & \multicolumn{1}{c}{$a_1$} & \multicolumn{1}{c}{$a_2$} & \multicolumn{1}{c}{$a_3$} & \multicolumn{1}{c}{$a_4$} & \multicolumn{1}{c}{$a_5$} & \multicolumn{1}{c}{$b$} & \multicolumn{1}{c}{$c$} & \multicolumn{1}{c}{$d_1$}
          & \multicolumn{1}{c}{$d_2$}& \multicolumn{1}{c}{$e_1$}
          & \multicolumn{1}{c}{$e_2$}& \multicolumn{1}{c}{$f_1$}
          & \multicolumn{1}{c}{$f_2$} & \multicolumn{1}{c}{$g$}
          \\
\hline
\multirow{1}{*}{p} & 94.1 & $-$831  & 0  & 16700   & $-$10200   & 0  & 10800 & 8590 & $-$4230000 & 3190 & 274000 & 17.4 & $-$39400 & 0.464 & 0  \\
\multirow{1}{*}{He} & 1.14 & 0 & $-$118 & 578 & 0 & $-$87   & 3120   &$-$5530 & 3370 & 1.29 & 134000 & 88.5 & $-$1170000 & 861 & 0.03 \\
\end{tabular}
\end{ruledtabular}
\end{table*}

\begin{figure}[!tbp]
\includegraphics[width=\linewidth]{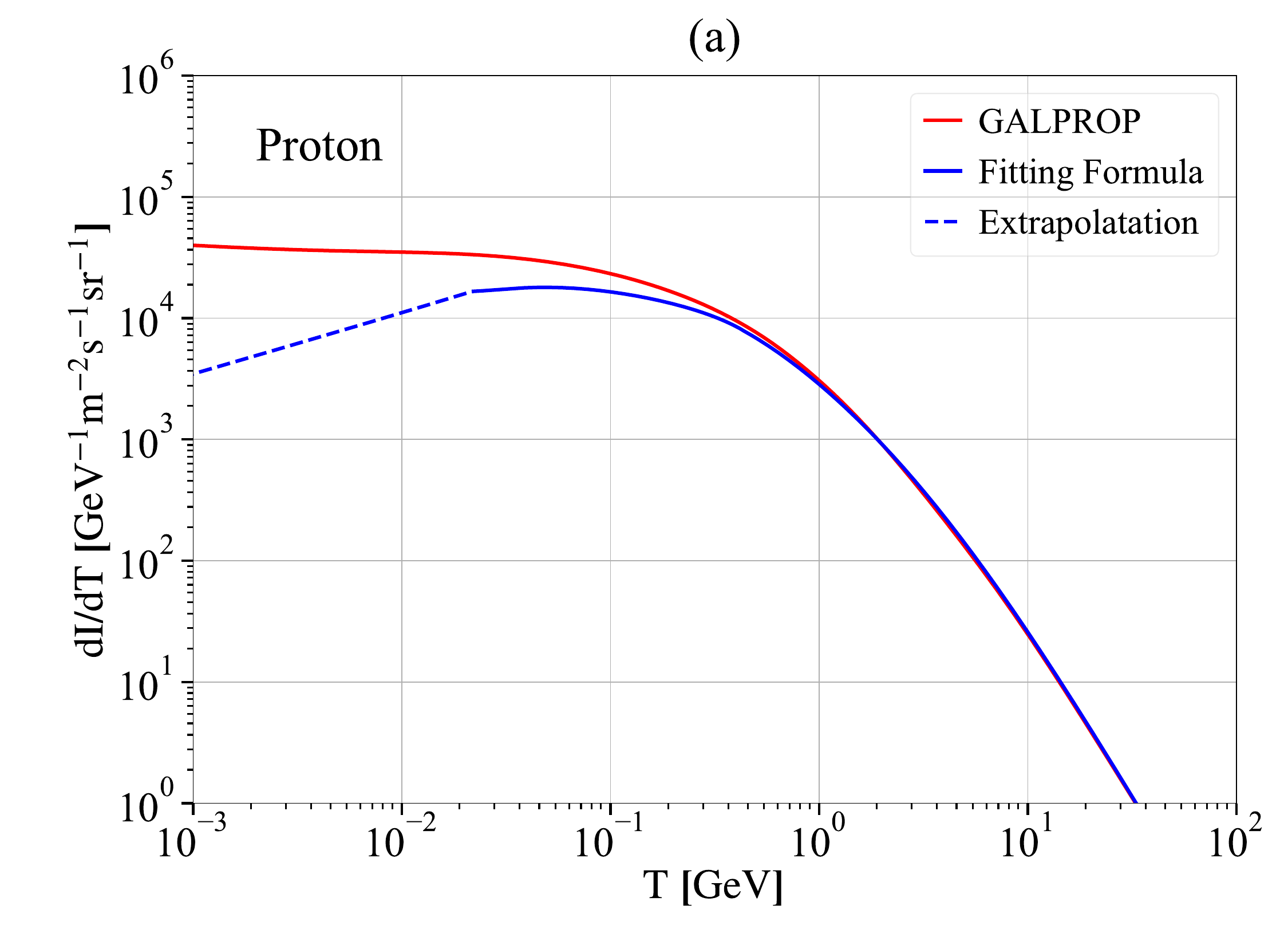}
\includegraphics[width=\linewidth]{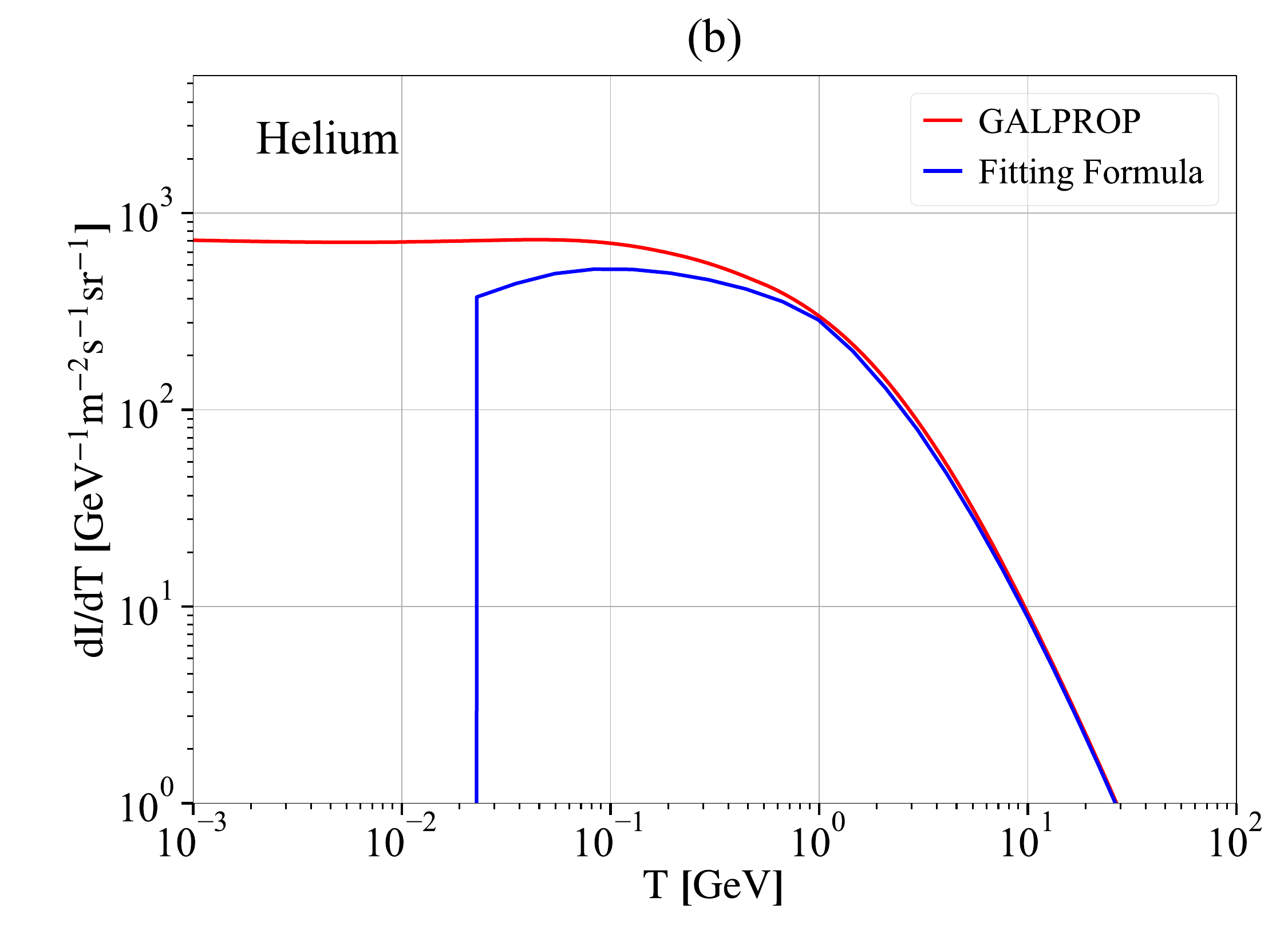}
\caption{
The local differential intensities of (a) proton and (b) helium, generated using the $\tt GALPROP$ simulation package, are shown as red lines. The intensities calculated using the fitting formula presented in Ref.~\cite{boschini_2017} are shown as blue lines. The dashed line represents the extrapolation in Ref.~\cite{CRDM_prl}.
}
\label{fig::cr}
\end{figure}

However, in the CRDM scenario, the DM particles are boosted by CRs not only near the Solar System but in the entire Milky Way, on which scale the solar modulation effect is negligible. The unmodulated CR flux data should be used when computing the CRDM flux. $\tt GALPROP$~\cite{boschini_2017,Boschini_2018} is a simulation package that uses real Milky Way information including the gas distribution and magnetic field distribution to model the CR diffusion and propagation in the Galaxy. In $\tt GALPROP$, the solar modulation is not considered, making it more appropriate for the CRDM analysis. In this analysis, the $\tt GALPROP$ package is used to generate the required CR data, including the spatial distribution and the flux of CR species from hydrogen to nickel in the Galaxy. The local differential intensities of proton and helium generated by $\tt GALPROP$ are shown in Fig.~\ref{fig::cr}. For comparison, the calculation results obtained using the fitting formula are also superimposed, which are slightly lower than those generated by $\tt GALPROP$ in the low-energy range.

\begin{figure}[!htbp]
\includegraphics[width=\linewidth]{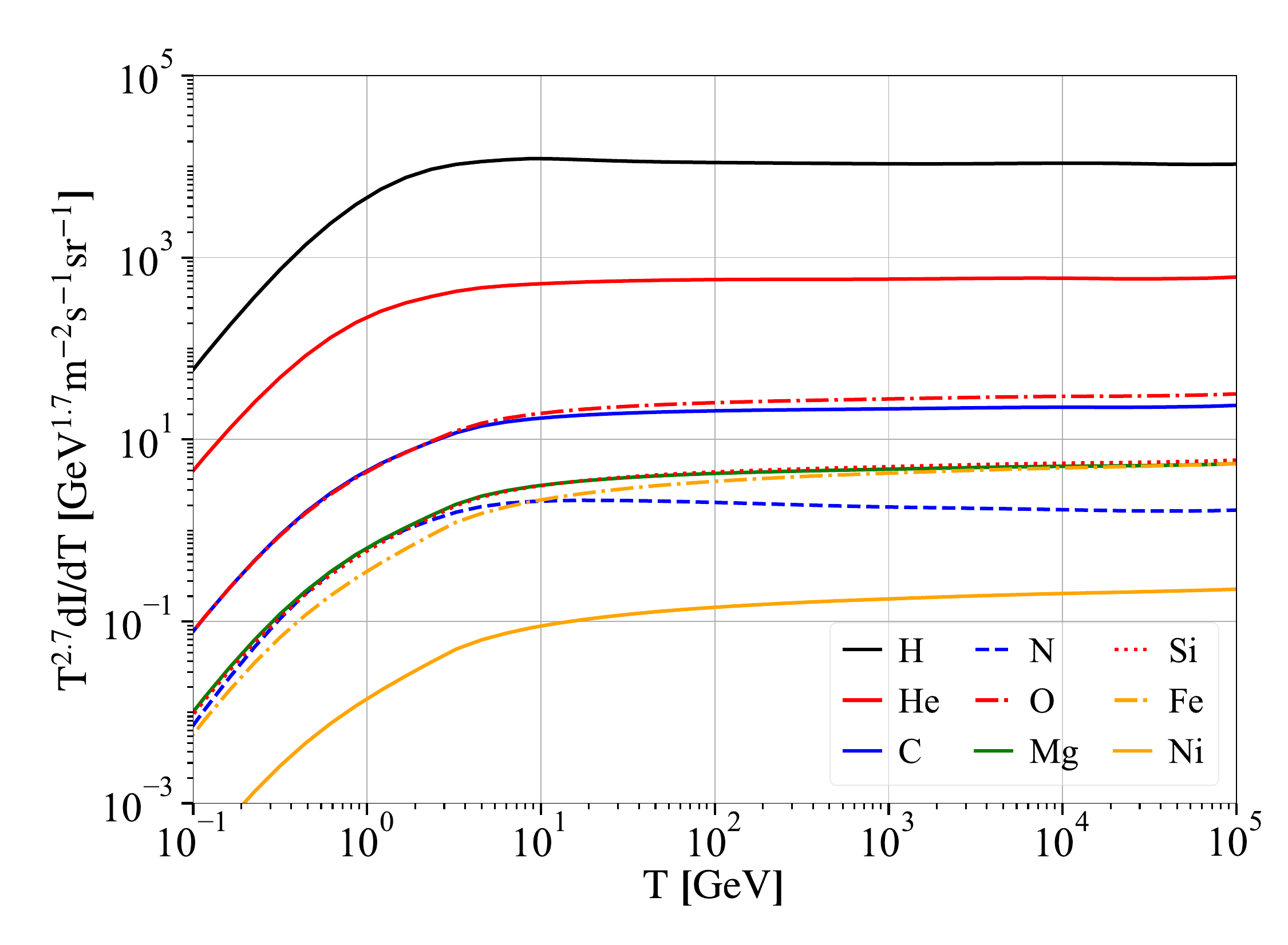}
\caption{
The local differential intensities scaled by $T^{2.7}$ of CR species H, He, C, N, O, Mg, Si, Fe, and Ni, generated using $\tt GALPROP$. $T$ denotes the kinetic energy. 
}
\label{fig::species}
\end{figure}

During the calculation of CRDM flux, the nuclei of H, He, C, N, O, Mg, Si, Fe, and Ni in the cosmic rays are included. Figure~\ref{fig::species} shows the local differential intensities of these CR species. Electrons are not considered here, as our main focus is $\chi$-N scattering.

\subsection{\label{sec_CRDM}CRDM calculation}
The calculation algorithm of the CRDM flux follows the steps presented in previous studies~\cite{CRDM_prl,CRDM_neutrino,CRDM_PRD,CRDM_TJ,CRDM_ZYF}. Compared with the relativistic CR particles, DM particles with velocities of $\sim 10^{-3}\ c$ can be effectively treated at rest. The kinetic energy transferred to a DM particle by an incident CR particle $i$ ($i$ = H, He,...) with mass
$m_i$ and kinetic energy $T_i$ is given by
\begin{equation} \label{eq2}
\begin{aligned}
T_{\chi}&=T^{max}_{\chi}\frac{1-\rm cos{\theta}}{2},\\
T_{\chi}^{max}&=\frac{T_i^2+2m_iT_i}{T_i+(m_i+m_{\chi})^2/(2m_{\chi})}.
\end{aligned}
\end{equation}

Here, $\theta$ represents the scattering angle in the center-of-mass system. Inverting Eq.~\ref{eq2} indicates the minimal incident energy required to produce $T_{\chi}$,
\begin{equation} \label{invert}
  T_i^{min}=\left(\frac{T_{\chi}}{2}-m_i\right)\left(1\pm\sqrt{1+\frac{2T_{\chi}}{m_{\chi}}\frac{(m_i+m_{\chi})^2}{(2m_i-T_{\chi})^2}}\right) , 
\end{equation}
where $+(-)$ corresponds to the case $T_{\chi}>2m_i$ ($T_{\chi}<2m_i$).

The CR nucleus-DM scattering differential cross section is given by
\begin{equation} \label{form}
 \frac{d\sigma_{\chi i}}{dT_{\chi}}=\frac{\sigma_{\chi N}}{T_{\chi}^{max}}A_i^2(\frac{\mu_{\chi i}}{\mu_{\chi N}})^2G_i(Q^2),    
\end{equation}
where $\sigma_{\chi N}$ represents the zero-momentum transferred DM-nucleon cross section, $A_i$ denotes the mass number of CR species $i$, $\mu_{\chi i}$ represents the DM-nucleus reduced mass, $\mu_{\chi N}$ denotes the DM-nucleon reduced mass, $G$ represents the form factor, which is related to the momentum transfer, and $Q=\sqrt{2m_{\chi}T_{\chi}}$. For proton and helium, we adopt the dipole form factor $G_i(Q^2)=1/(1+Q^2/\Lambda_i^2)^2$~\cite{dipole}, where the $\Lambda_p\simeq770$ and $\Lambda_{He}\simeq410$ MeV~\cite{p_he_form}. For other heavier nuclei, we adopt the conventional Helm form factor~\cite{formfactor2}.

The CR-induced DM flux is obtained by integrating over all CR nucleus species $i$ and energies $T_i$ along the line of sight (LOS),
\begin{equation} \label{eq:crdm}
\begin{aligned}
\frac{d\Phi_{\chi}}{dT_{\chi}}&=\sum_i\int\frac{d\Omega}{4\pi}\int_{LOS} dl \int_{T_{i}^{min}}^{\infty}\frac{\rho_{\chi}(\boldsymbol{r})}{m_{\chi}}\frac{d\sigma_{\chi i}}{dT_{\chi}}\frac{d\Phi_i(\boldsymbol{r})}{dT_i}dT_i,
\end{aligned}
\end{equation}
where $\Omega$ represents the solid angle and $T_{i}^{min}$ denotes the minimal CR energy required to produce a DM recoil energy $T_{\chi}$, which can be calculated by Eq.~\ref{invert}. For the DM density distribution $\rho_{\chi}(\boldsymbol{r})$, we adopt the Navarro-Frenk-White (NFW) profile~\cite{nfw}. The spatial distribution of the flux of CR species $i$, $\frac{d\Phi_i(\boldsymbol{r})}{dT_i}$, is related to the distance to the Galactic Center $\boldsymbol{r}$.

\begin{figure}[!htbp]
\includegraphics[width=\linewidth]{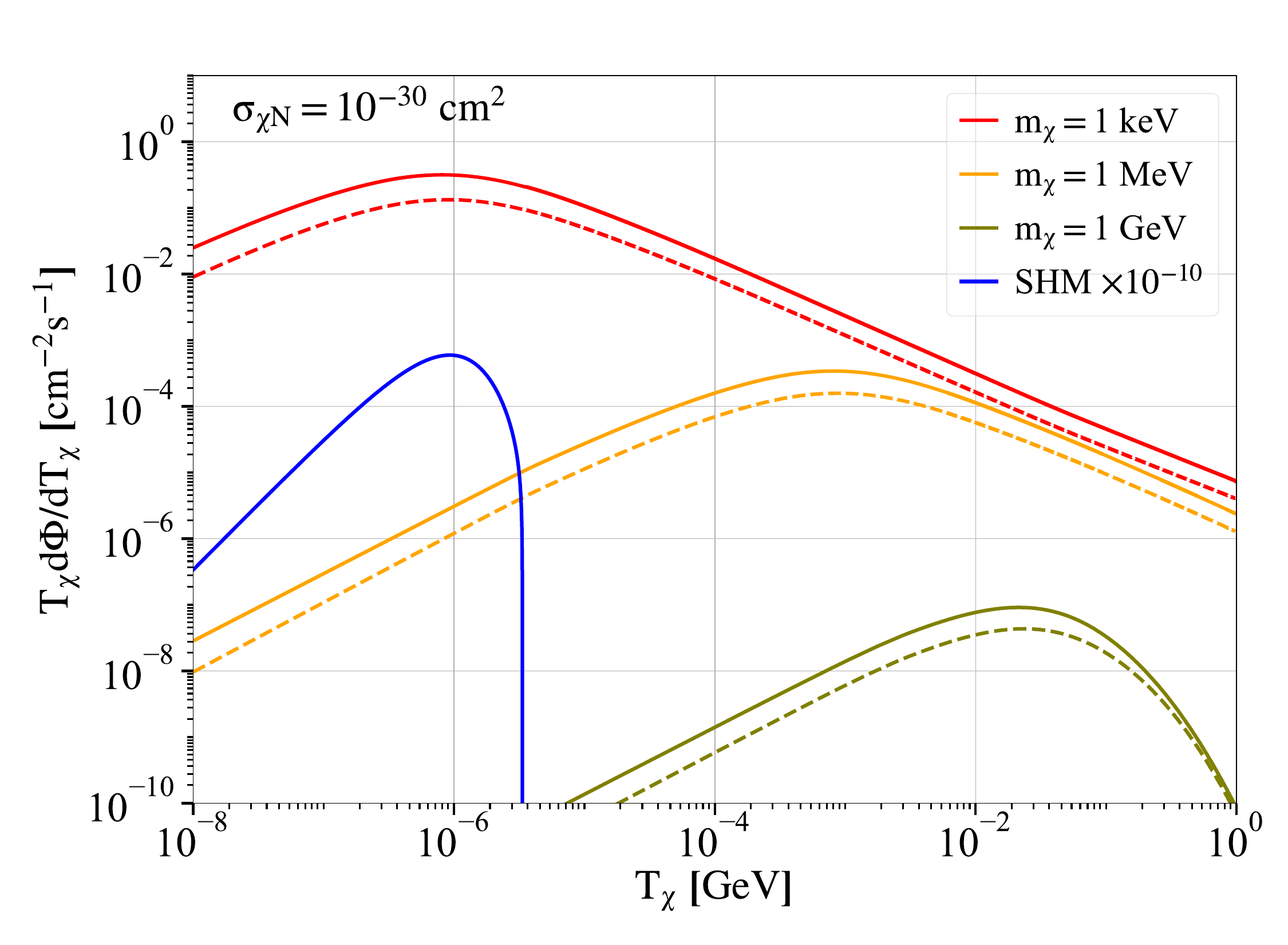}
\caption{
The CRDM fluxes scaled by $T_{\chi}$ for different DM masses with $\sigma_{\chi N} = 10^{-30}\ \rm cm^2$. The solid lines indicate the total flux contributed by all the considered CR species, while the dashed lines represent those contributed by the proton and helium only. The flux of DM with 1~GeV mass derived from the SHM is rescaled by a factor $10^{-10}$ for comparison.
}
\label{fig::crdm}
\end{figure}

Using Eq.~\ref{eq:crdm} and the CR data generated by $\tt GALPROP$, the CRDM fluxes for various DM masses are obtained and shown in Fig.~\ref{fig::crdm}. The dominant contributions are from proton and helium, while the total contribution from other heavier CR species accounts for approximately half of the total CRDM flux. The flux of DM with 1 GeV mass derived from the SHM and rescaled by a factor $10^{-10}$ is also shown for comparison. It can be inferred that a very tiny amount of DM particles are accelerated to extremely high energy. This work only considers CRDM particles with kinetic energies below 1 GeV to avoid possible inelastic processes~\cite{CRDM_prl}.

\subsection{\label{sec_Deff}Discussion on the effective distance}
For simplicity, an effective distance $D_{eff}$ is introduced to represent the space integral in Eq.~\ref{eq:crdm},
\begin{equation} \label{eq:crdm2}
\begin{aligned}
\frac{d\Phi_{\chi}}{dT_{\chi}}=\frac{\rho_{\chi}^{local}}{m_{\chi}}D_{eff}\sum_i\int_{T_i^{min}}^{\infty} \frac{d\sigma_{\chi i}}{dT_{\chi}}\frac{d\Phi_i^{LIS}}{dT_i}dT_i,
\end{aligned}
\end{equation}
where the local DM density $\rho_{\chi}^{local}=0.3\ \rm GeV/cm^3$~\cite{nfw} and $\frac{d\Phi_i^{LIS}}{dT_i}$ is the Local Interstellar Spectrum (LIS) of CR flux. The expression of $D_{eff}$ can be obtained by comparing Eqs.~\ref{eq:crdm} and~\ref{eq:crdm2}; their comparison provides the rigorous definition of the spatial coefficient $D_{eff}$,
\begin{equation} \label{eq:deff}
\begin{aligned}
D_{eff}(T_{\chi})=\frac{\sum_i\int\frac{d\Omega}{4\pi}\int_{LOS} dl \int_{T_{i}^{min}}^{\infty}\frac{\rho_{\chi}(\boldsymbol{r})}{m_{\chi}}\frac{d\sigma_{\chi i}}{dT_{\chi}}\frac{d\Phi_i(\boldsymbol{r})}{dT_i}dT_i}{\sum_i\int_{T_i^{min}}^{\infty} \frac{d\sigma_{\chi i}}{dT_{\chi}}\frac{d\Phi_i^{LIS}}{dT_i}dT_i}.
\end{aligned}
\end{equation}
 
\begin{figure}[!htbp]
\centering
\includegraphics[width=\linewidth]{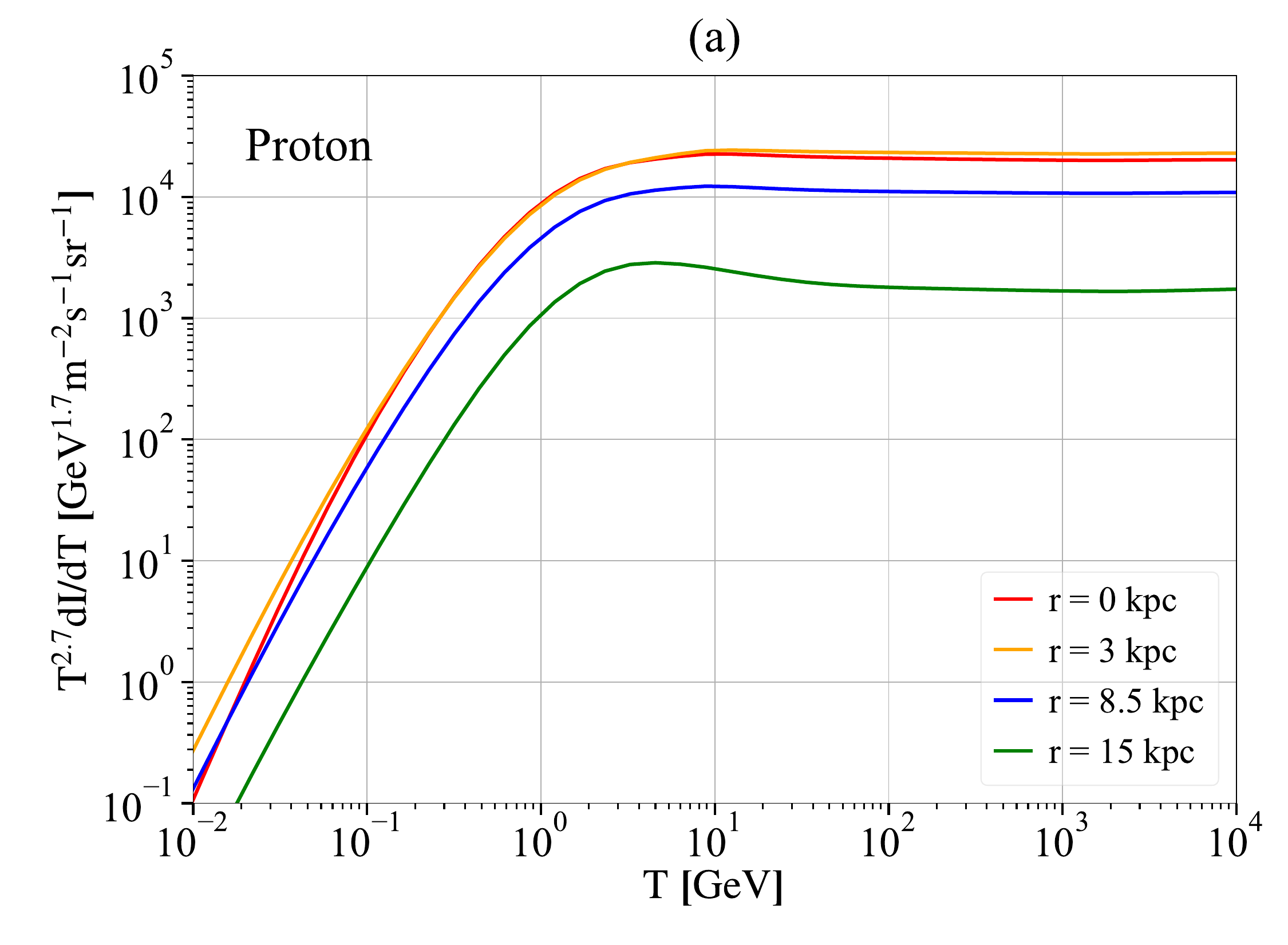}
\includegraphics[width=\linewidth]{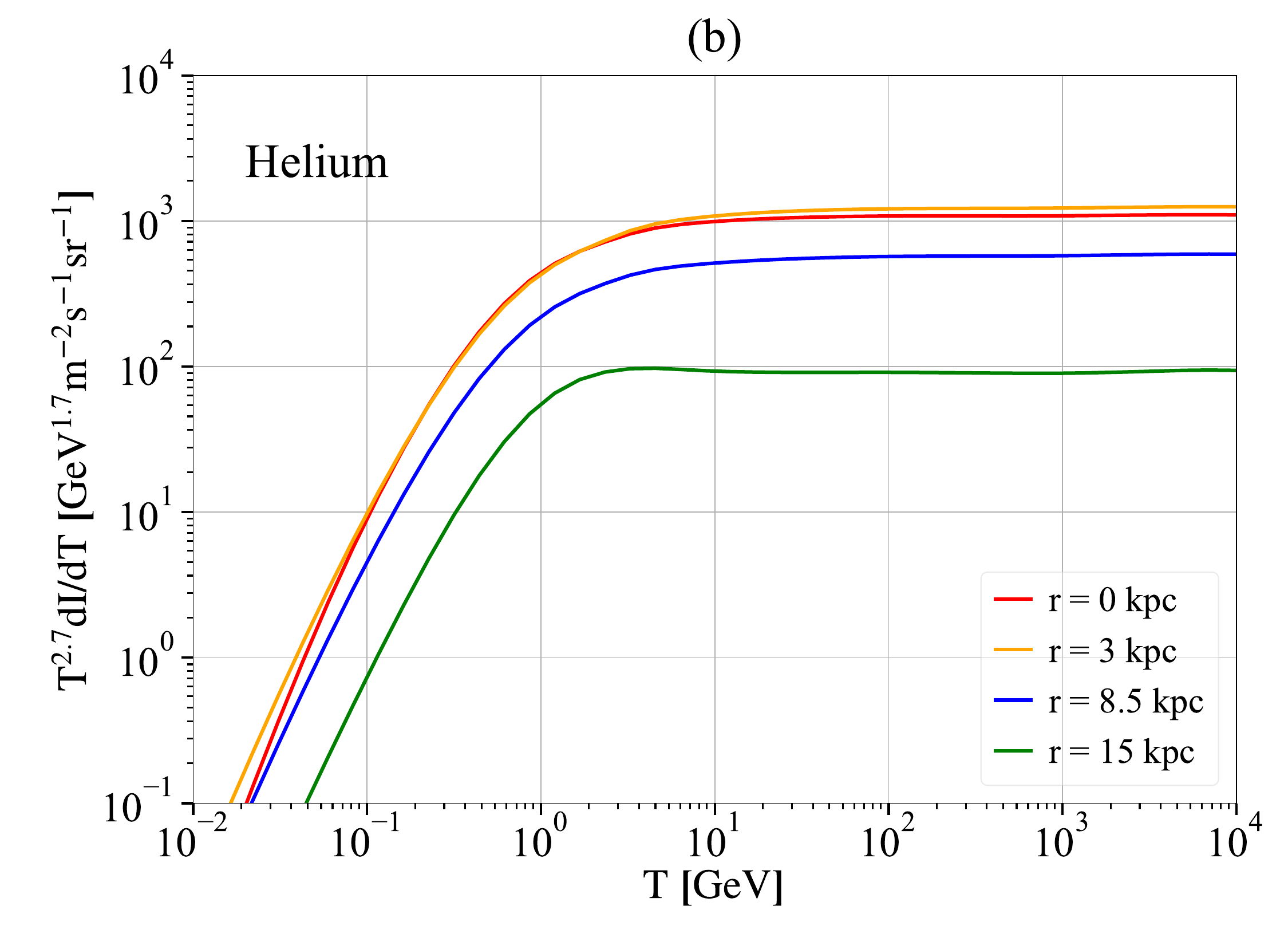}
\includegraphics[width=\linewidth]{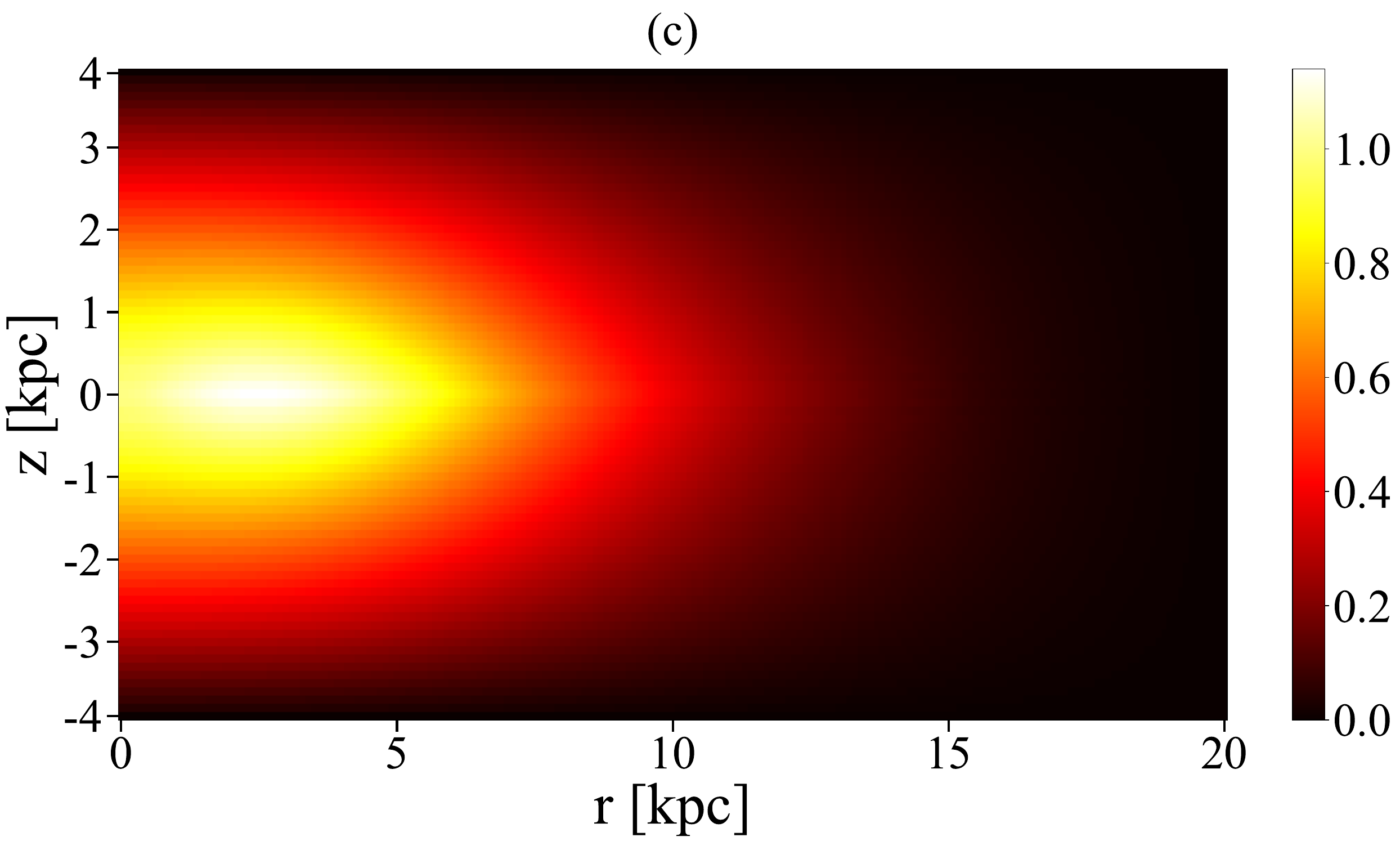}
\caption{
The CR (a) proton and (b) helium intensities as a function of kinetic energy $T$ in the Galactic plane ($z=0$) at different distances to the Galactic Center. (c) The spatial distribution of CR proton intensity in the Galaxy. The proton intensity in the Galactic Center ($r=$ 0, $z =$ 0) is set to be 1. 
}
\label{fig::dis}
\end{figure}

Depending on the DM kinetic energy $T_{\chi}$ and the DM mass $m_{\chi}$, the calculation of $D_{eff}$ requires the CR spatial flux and the range of the spatial integral as inputs. Regarding the spatial distribution of the flux of CR species $i$, $\frac{d\Phi_i(\boldsymbol{r})}{dT_i}$, simulation results from $\tt GALPROP$ show that the overall CR flux in the Galactic Center is several times higher than the flux near the Solar System, while their spectral shapes remain similar. Figure~\ref{fig::dis} demonstrates the inhomogeneity of the proton and helium intensities in the Galaxy. The proton flux at the Galactic Center is 2--3 times larger than the LIS ($r$ = 8.5 kpc, the location of the Solar System) and one magnitude larger than the flux at $r$ = 15 kpc.

Conventionally, the CR distribution was treated as homogeneous, under which assumption the factor related to $T_{\chi}$ in Eq.~\ref{eq:deff} will be canceled out, leading to $D_{eff}=\int \frac{d\Omega}{4\pi} \int \frac{\rho_{\chi} (\boldsymbol{r})}{\rho_{\chi}^{local}} dl$. The NFW  profile~\cite{nfw} is used to set DM density $\rho_{\chi} (\boldsymbol{r})$ and $\rho_{\chi}^{local}$. The value of $D_{eff}$ is determined by the integral range selected in this case. 

Different CR spatial distribution models have been used in previous studies. A spherical CR distribution assumption and a full line-of-sight integration to radius $r$ = 1 kpc (10 kpc) resulting in $D_{eff}$ = 0.997 kpc ($D_{eff}$ = 8.02 kpc) were adopted for XENON-1T data~\cite{CRDM_prl}, where the $D_{eff}$ values of 1 and 10 kpc were used as benchmark values. Similarly, a cylindrical distribution with a radius of 10 kpc and half-height of 1 kpc was used for KamLAND data~\cite{CRDM_PRD}, providing $D_{eff}$ = 3.7 kpc. We note that taking a larger halo (such as a cylinder with a half-height of 4--10 kpc) could be more realistic according to astronomical studies~\cite{Strong_2000}.

\begin{figure}[!tbp]
\includegraphics[width=\linewidth]{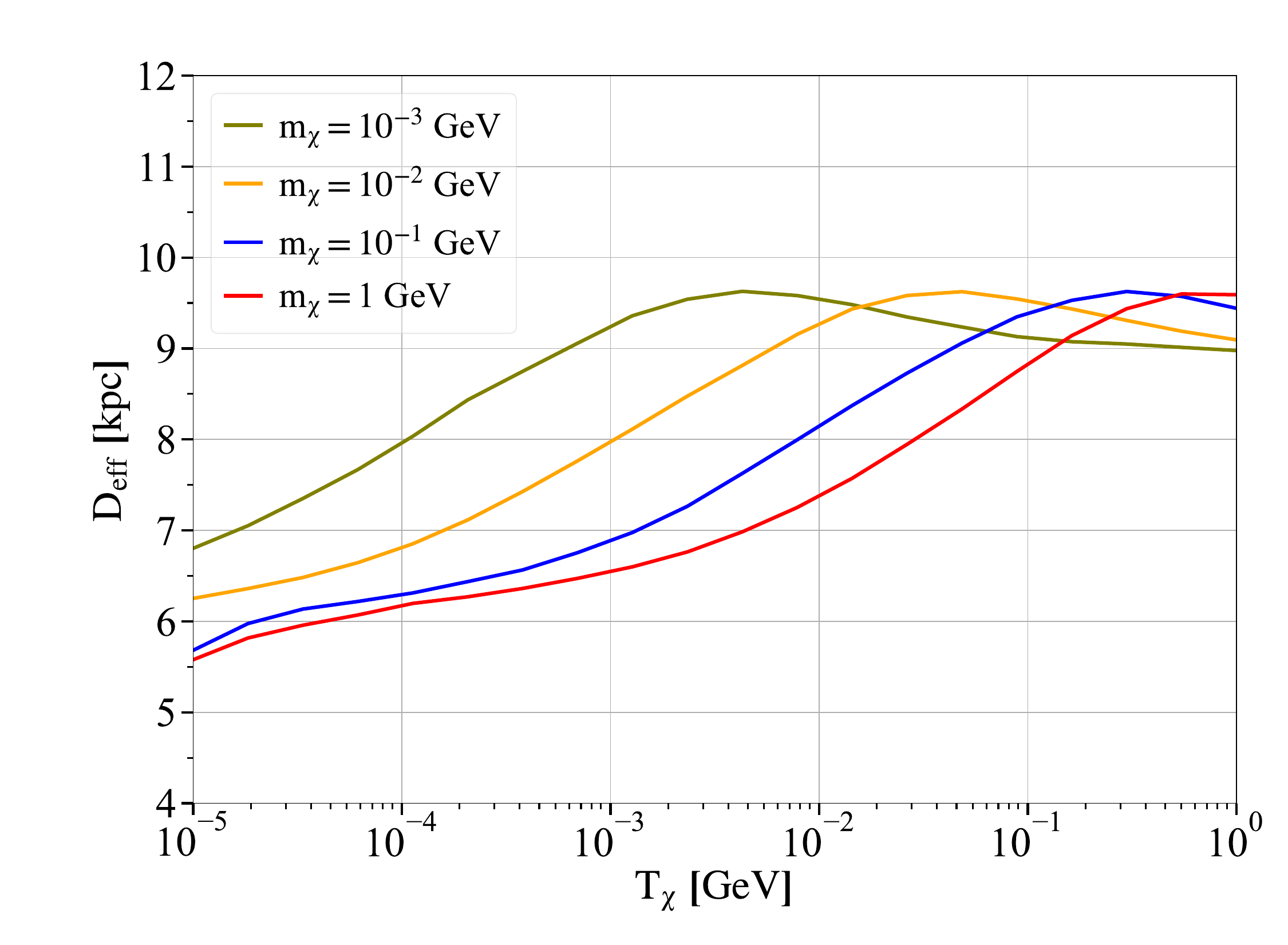}
\caption{
The effective distance $D_{eff}$ as a function of DM kinetic energy with different DM masses from 1 MeV to 1 GeV. 
}
\label{fig::deff}
\end{figure}

In this analysis, the CR spatial distribution data are generated using $\tt GALPROP$. Further, a cylindrical distribution of CR with radius $r=20$ kpc and half-height $z=4$ kpc is adopted. The $D_{eff}$ values as a function of DM kinetic energy of different DM masses are calculated using Eq.~\ref{eq:deff} and shown in Fig.~\ref{fig::deff}. The $D_{eff}$ values start from about 6 kpc at low energies, reach peaks at about 10 kpc, and then decrease slightly. As the $D_{eff}$ values used in previous studies, the 1~\cite{CRDM_prl} and 3.7 kpc~\cite{CRDM_PRD} are too conservative, while 10 kpc~\cite{CRDM_prl} is a slightly overestimated value. The value of $D_{eff}$ is a major source of uncertainty in the CRDM analyses. To avoid the uncertainty related to $D_{eff}$, we do not adopt a fixed $D_{eff}$ value in this analysis. The CRDM fluxes are calculated using Eq.~\ref{eq:crdm}, which is more accurate.

\section{\label{sec3}Earth attenuation and recoil spectrum}
The DM particles must travel through a few kilometers of rock before reaching the underground laboratory. The DM flux will be changed from $d\Phi_{\chi}/dT_{\chi}$ to $d\Phi_{\chi}/dT^z_{\chi}$, where $z$ represents the depth of rock through which incident DM particles pass. 
 
A Monte Carlo simulation package $\tt CJPL\_ESS$~\cite{cjpless} was developed to simulate the Earth shielding effect of CJPL, encompassing a detailed geometric model and the rock compositions of Jinping Mountain. In the $\tt CJPL\_ESS$ simulation framework, the initial DM particles of different masses are generated on a sphere concentric with Earth with a radius of $R_0 = R_{\oplus}+h$, where Earth's radius $R_{\oplus}$ is 6400 km and $h$ = 3 km. The kinetic energies of the initial DM particles are sampled according to the CRDM flux. Both the energy loss and the direction variation of a DM particle after each $\chi$-N scattering in Earth's crust are considered until it arrives at the underground laboratory after multiple scatterings. The number and energies of the DM particles arriving at the underground laboratory are counted to reconstruct the attenuated CRDM flux $\frac{d\Phi_{\chi}}{dT^z_{\chi}}$.

For relativistic CRDM particles with considerable kinetic energy, the momentum transfers in the $\chi$-N scattering process can be large; thus, the impact of the nuclear form factor of nuclei in Earth's crust cannot be neglected while it was treated as unity for simplicity in some previous analysis~\cite{CRDM_prl,CRDM_TJ,CRDM_ZYF}. The form factor suppression of the cross section results in a larger mean free path and lower collision rates. Furthermore, the energy losses of DM particles are reduced because the cross sections become smaller for large momentum transfer scenarios. The attenuated CRDM flux can extend to a considerably higher energy than that in the case without considering the form factor, which significantly improves the sensitivity of the DD experiments to larger cross sections~\cite{es_zyf}. We consider the effect of the form factor here in the $\tt CJPL\_ESS$ simulation package~\cite{cjpless}. The free-propagation length between two scatterings is sampled using the mean free path calculated via the total cross section obtained by integrating Eq.~\ref{form}. The momentum transfer $Q$ of each scattering is sampled according to the differential cross section corrected by the form factor $G(Q^2)$. The Helm form factor is used here for elements in  Earth's crust~\cite{formfactor2}.

The differential event rate of the $\chi$-N elastic scattering in the detectors is calculated using the following equation:
\begin{equation}
\frac{dR}{dE_R}=N_TA^2 (\frac{\mu_{\chi A}}{\mu_{\chi N}})^2\int_{T^{min}_{\chi}}\frac{\sigma_{\chi N}}{E_{R}^{max}} G_A^2(Q^2)\frac{d\Phi_{\chi}}{dT^z_{\chi}}dT^z_{\chi},
\end{equation}
where $E_R$ is the nuclear recoil energy, $N_T$ is the number of target nuclei per unit detector mass, $A$ is the mass number of Ge nucleus, and $\mu_{\chi A}$ is the DM-Ge nucleus reduced mass. The value of $E_{R}^{max}$ is obtained using Eq.~\ref{eq2} by replacing $i\rightarrow\chi$ and $\chi \rightarrow N$. Inverting the expression of $E_{R}^{max}$ affords $T^{min}_{\chi}$. $G_A(Q^2)$ is the nuclear form factor, for which the Helm form factor~\cite{formfactor2} is used.

In a germanium semiconductor detector, the observed total deposit energy $E_{det}$ is different from the real nuclear recoil energy $E_R$ and should be corrected by the quenching factor, $E_{det}=Q_{nr}E_R$. The quenching factor in Ge is calculated using the Lindhard formula~\cite{lindhard} ($\kappa = 0.16$, a typical value reported in the literature and well matches recent measurements at low-energy range~\cite{quenching2022}) with a 10\% systematic error adopted in this analysis. 

\begin{figure}[!htbp]
\includegraphics[width=\linewidth]{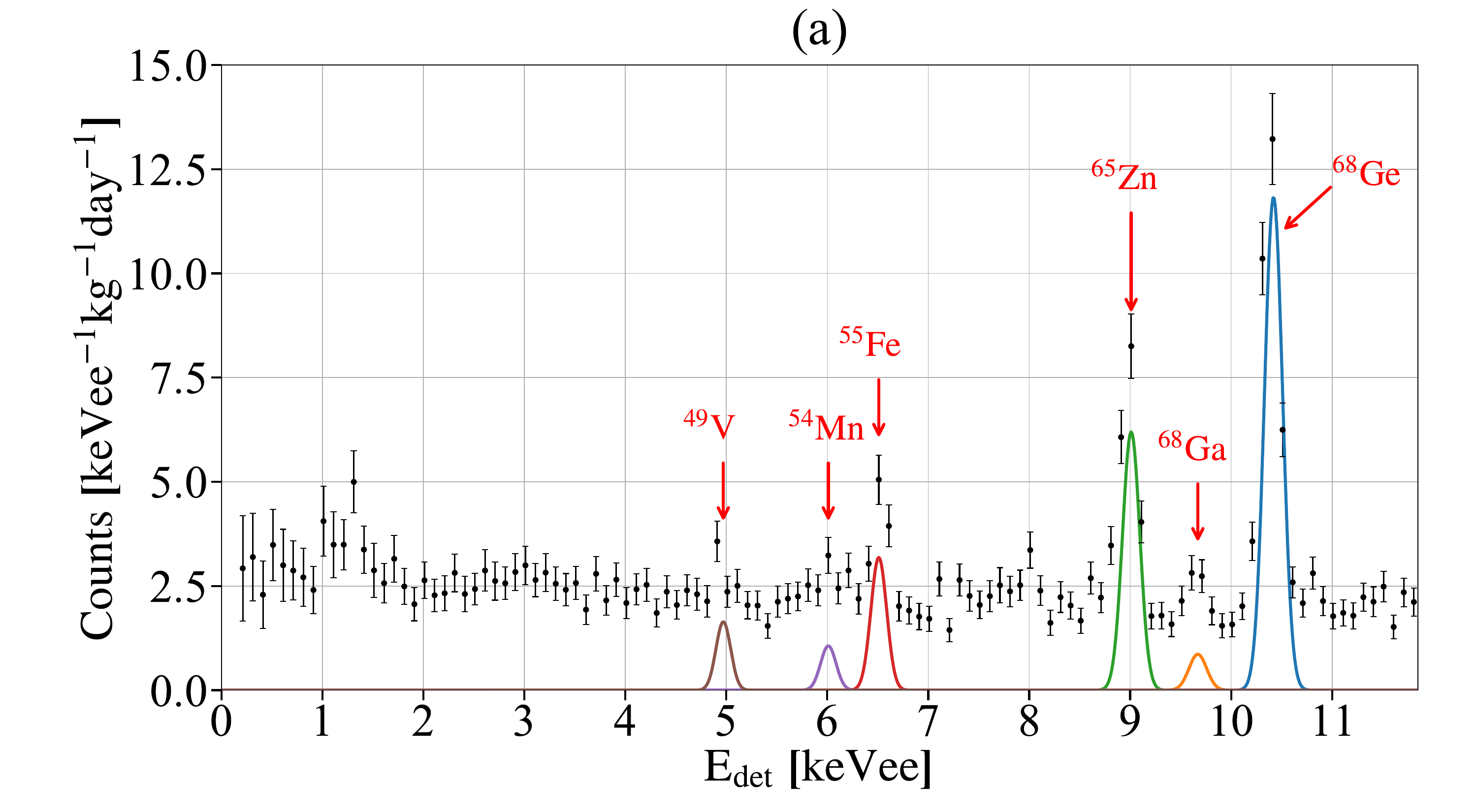}
\includegraphics[width=\linewidth]{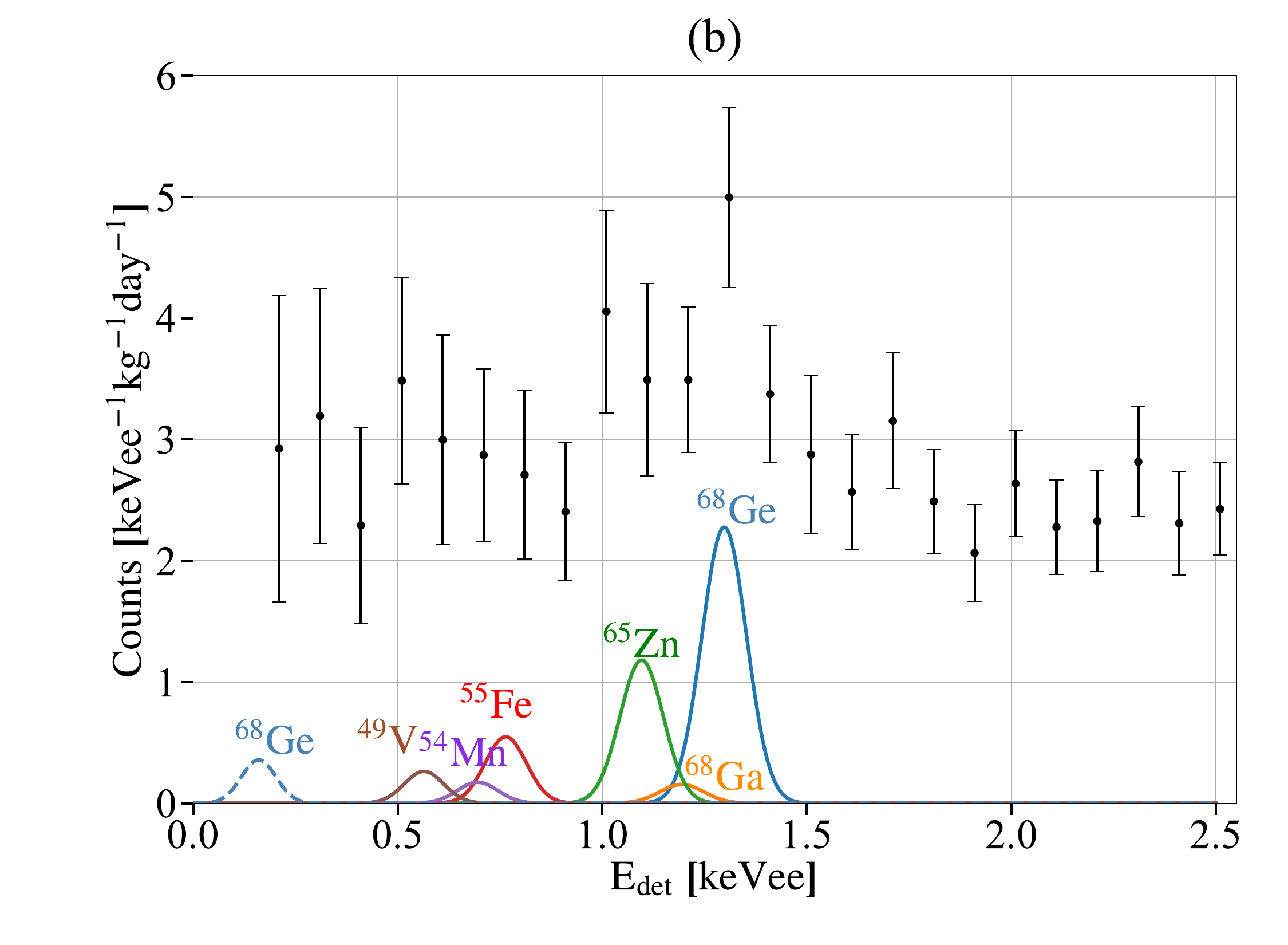}
\includegraphics[width=\linewidth]{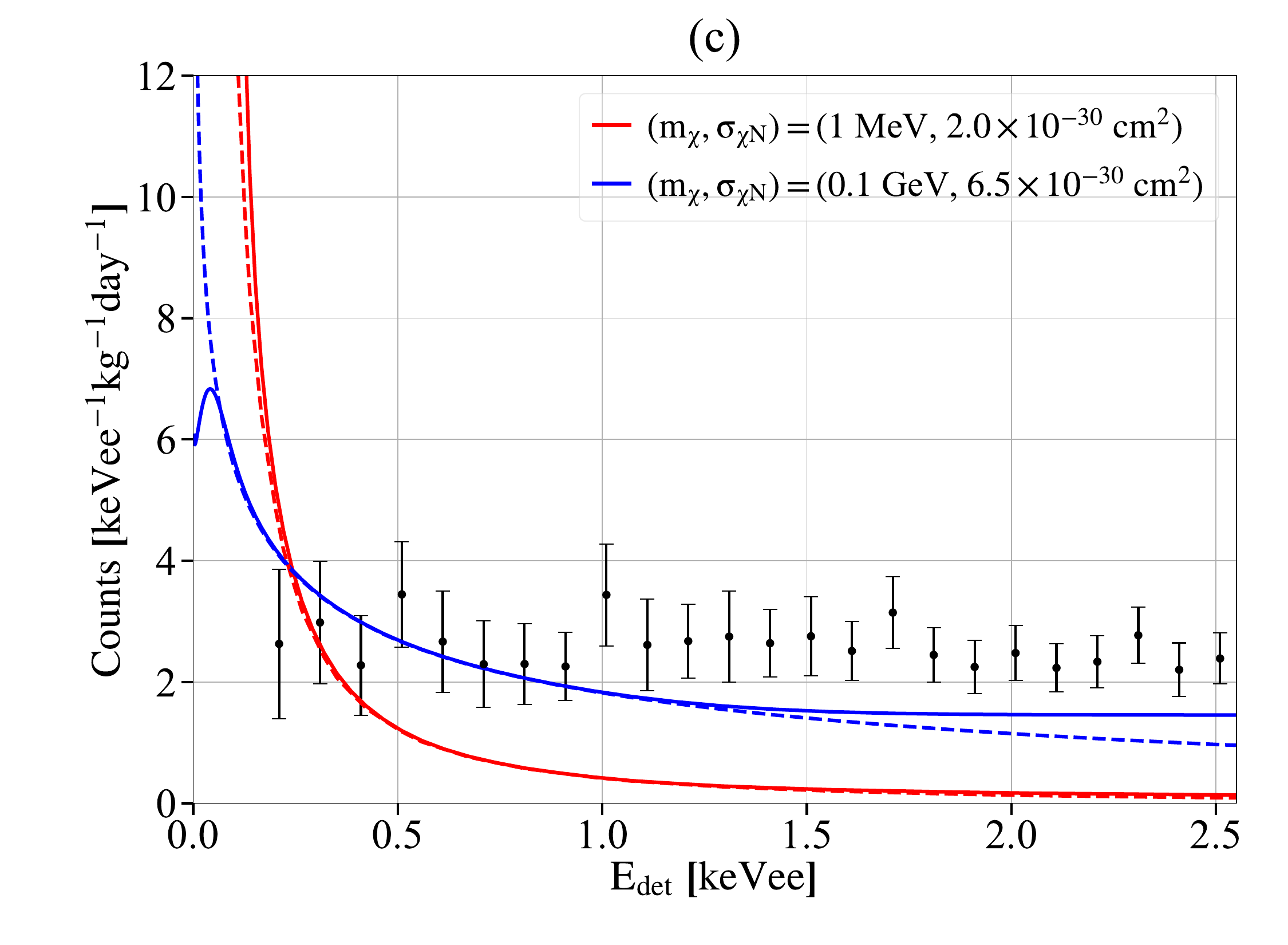}
\caption{
(a) The measured spectrum with error bars based on the 205.4 kg day dataset of the CDEX-10 experiment. The bin width is 100 eVee and the energy range is 0.16--12 keVee. The arrows show the characteristic K-shell x-ray peaks from internal cosmogenic radionuclides, including $\rm ^{68}Ge$, $\rm ^{68}Ga$, $\rm ^{65}Zn$, $\rm ^{55}Fe$, $\rm ^{54}Mn$, and $\rm ^{49}V$. The contributions of these radionuclides derived by the best fit of the spectrum are superimposed.
(b) The contributions of L- and M-shell x-ray peaks are derived from the corresponding K-shell line intensities~\cite{K-X-ray}. The L-shell x-ray peaks are shown in solid lines. The dashed line represents the M-shell x-ray peak of $\rm^{68}Ge$.
(c) Residual spectrum with the L- and M-shell x-ray contributions subtracted, together with the expected CRDM signals at (1 MeV, $2.0\times 10^{-30} \rm cm^2$) (red lines) and (0.1 GeV, $6.5\times 10^{-30} \rm cm^2$) (blue lines), which correspond to the lower bound of our exclusion region. Solid and dashed lines correspond to spectra with and without considering energy resolution, the standard deviation of which is 35.8 + 16.6$\times E^{\frac {1}{2}}$ (eV), where $E$ is expressed in keV.
}
\label{fig::spectrum}
\end{figure}

\section{\label{sec4}Exclusion results}
Data used in this CRDM analysis are from the CDEX-10 data taken from February 2017 to August 2018, with a total exposure of 205.4 kg day~\cite{cdex_darkphoton}, after considering the dead time corrections and the fiducial mass. The data analysis follows the procedures described in our earlier works, including energy calibration, physics event selection, bulk-surface event discrimination, and various efficiency corrections~\cite{cdex1b2018,cdex10_tech,cdex102018,cdex1b_am,cdex_darkphoton}. The spectrum obtained after a series of physics event selections and efficiency corrections is shown in Fig.~\ref{fig::spectrum}(a). The physics analysis threshold is 160 eVee, where the combined efficiency is 4.5\%~\cite{cdex10_tech}. The characteristic K-shell x-ray peaks from internal cosmogenic radionuclides like $\rm ^{68}Ge$, $\rm ^{68}Ga$, $\rm ^{65}Zn$, $\rm ^{55}Fe$, $\rm ^{54}Mn$, and $\rm ^{49}V$ can be identified. Their intensities are derived from the best fit of the spectrum~\cite{cdex102018}. The contributions of L- and M-shell x-ray peaks, which are not fitted, but are derived from the corresponding K-shell line intensities~\cite{K-X-ray}, are as demonstrated in Fig.~\ref{fig::spectrum}(b). The residual spectrum with the L- and M-shell x-ray contributions subtracted in the region of 0.16--2.56 keVee is shown in Fig.~\ref{fig::spectrum}(c). The expected CRDM spectra of DM mass 1 MeV, cross section $2.0\times 10^{-30} \rm cm^2$ and 0.1 GeV, cross section $6.5\times 10^{-30} \rm cm^2$ are also shown for comparison. As shown in Fig.~\ref{fig::spectrum}(c), light DM particles in the sub-GeV mass range can generate considerable signals above the detection threshold after the CR acceleration.

\begin{figure}[!htbp]
\includegraphics[width=\linewidth]{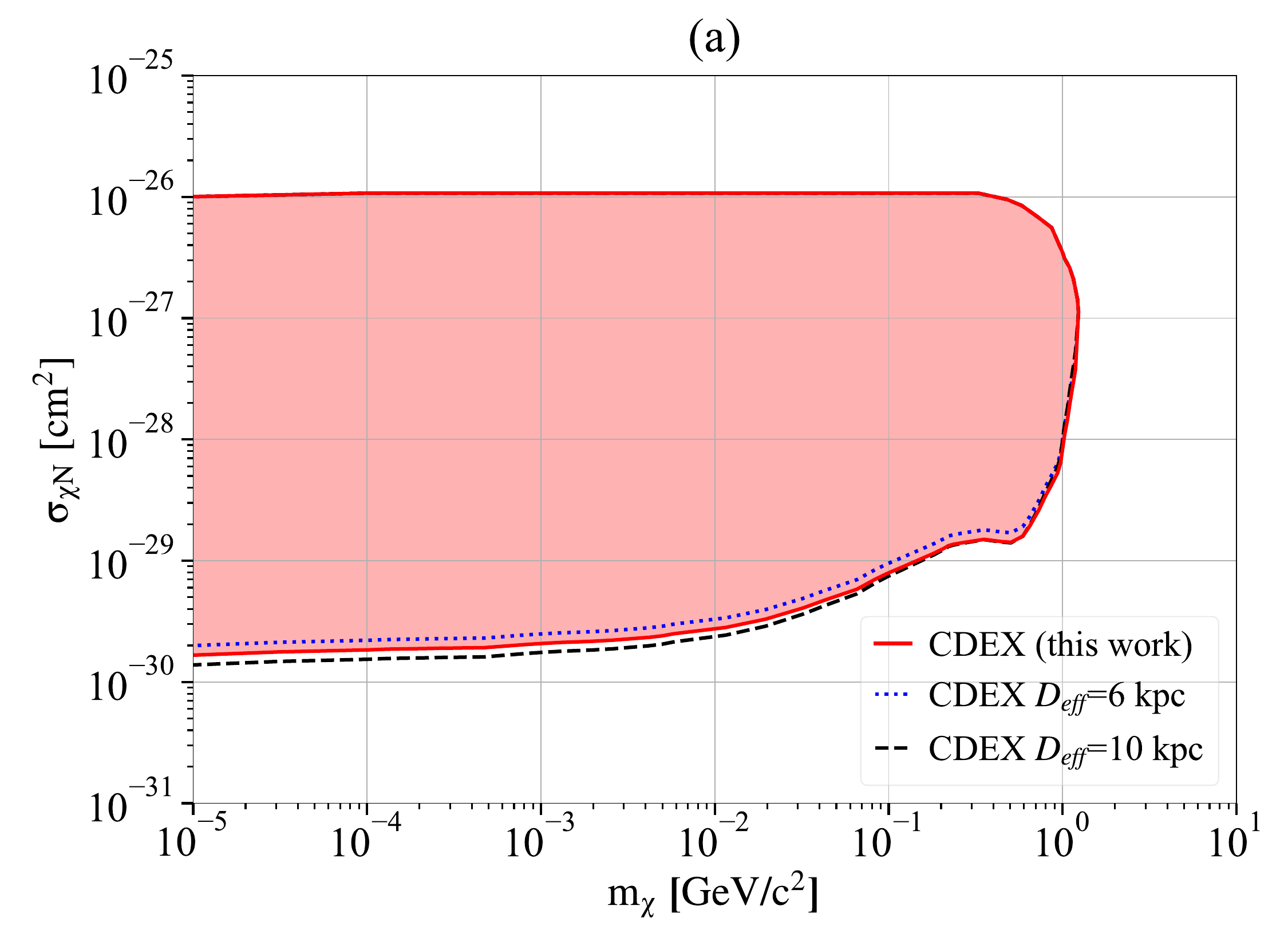}
\includegraphics[width=\linewidth]{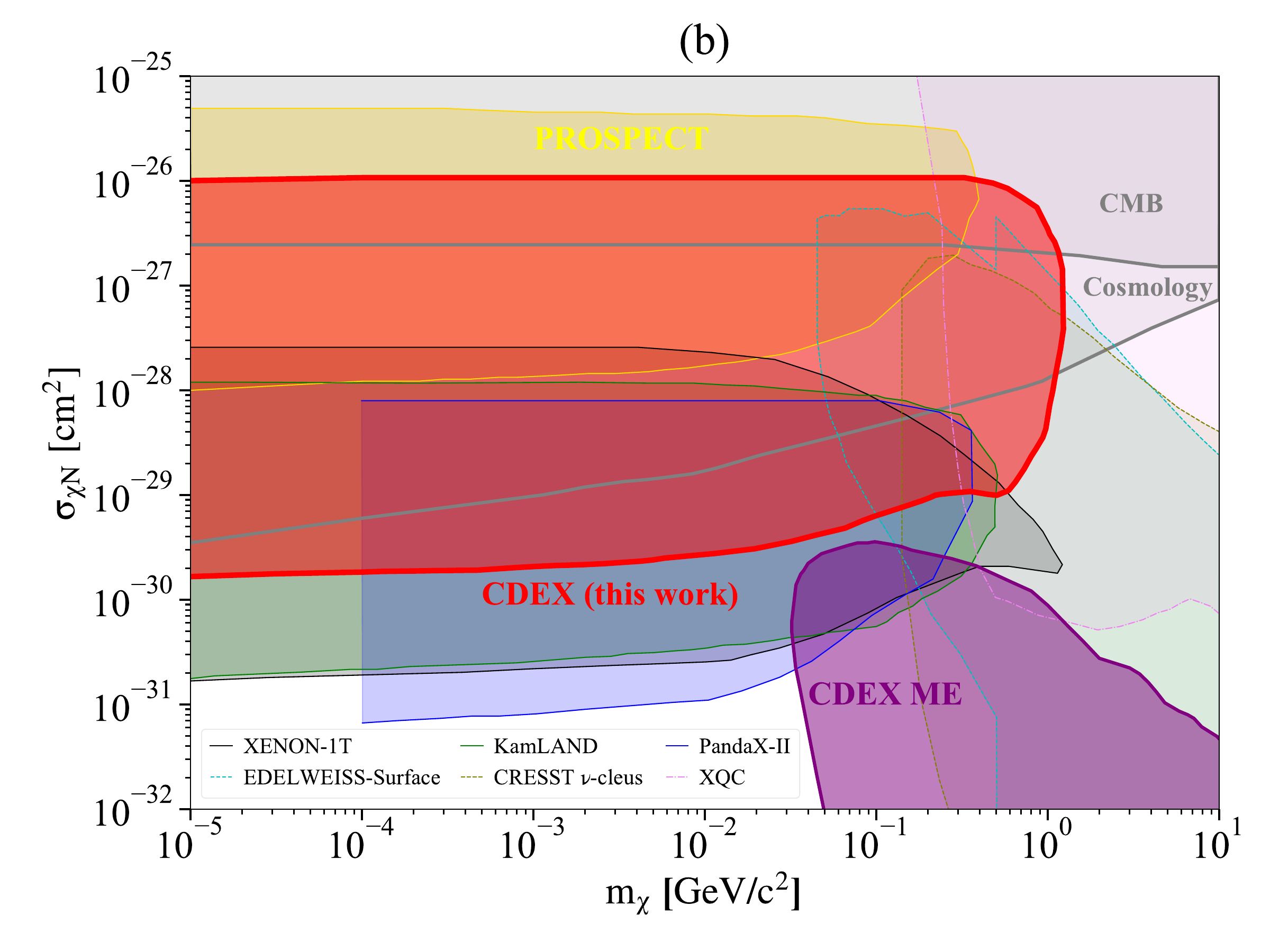}
\caption{
(a) The exclusion region derived from the 205.4 kg day dataset of the CDEX-10 experiment, represented in red solid contour, with the CR acceleration mechanism adopted. For comparison, the exclusion results derived from an effective distance $D_{eff}=$ 6 and 10 kpc are also presented. Since it (red curve) is more precise, applying a more accurate CR spatial distribution, we quote it as our official result.
(b) The comparison of our obtained result and the limits from previous studies. The exclusion region from PROSPECT and PandaX-II using CR acceleration~\cite{prospect,PandaX-CRDM} and the published limits under SHM scenario from CDEX-10 Migdal effect analysis (CDEX-ME)~\cite{cjpless}, CRESST $\rm\nu$-cleus 2017 surface run~\cite{CRESST-Surface,earthshielding2,cresstsurface}, EDELWEISS-Surface~\cite{edelweiss}, and the XQC experiment~\cite{XQC1,XQC2} are superimposed. 
The limits based on data from XENON-1T~\cite{CRDM_prl} and KamLAND~\cite{CRDM_PRD}, as well as selected cosmological constraints from CMB~\cite{cmb} and the large scale structure~\cite{cosmo11}, are superimposed. It should be noted that the lower bound of the KamLAND results in this plot was rescaled by a factor of 0.85 (i.e., an additional 17\% enhancement) relative to that of Ref.~\cite{CRDM_PRD}, due to misinterpretation of the ``13.5--20 MeV" energy bin of the KamLAND measured data~\cite{KamLAND}.
}
\label{fig::region}
\end{figure}

As shown in Fig.~\ref{fig::region}, the exclusion region of CRDM with a $90\%$ confidence level is derived using the binned Poisson method~\cite{binpoisson}. The Earth attenuation effect has been considered with the Monte Carlo simulation package $\tt CJPL\_ESS$ to derive the upper bound. Previous CRDM research carried out by PROSPECT and PandaX-II, and phenomenological interpretations using XENON-1T and KamLAND data are superimposed~\cite{prospect,PandaX-CRDM,CRDM_PRD,CRDM_prl}. Constraints from the CRESST $\rm\nu$-cleus 2017 surface run~\cite{CRESST-Surface,earthshielding2,cresstsurface}, EDELWEISS-Surface~\cite{edelweiss}, and X-ray Quantum Calorimeter experiment (XQC)~\cite{XQC2} are also shown in the figure. Limits from cosmological studies based on the CMB~\cite{cmb} and the large scale structure~\cite{cosmo11} are shown in gray contour. The constraints obtained from BBN presented in Ref.~\cite{bbn} are not included in the figure as the values obtained in Ref.~\cite{bbn} are model specific. With the form factor considered in the Earth attenuation calculations, this work excludes a large region from $1.7\times10^{-30}$ to $10^{-26}~\rm cm^2$ in the dark matter--nucleon elastic scattering cross section. This result corresponds to better sensitivities than the cosmological limits~\cite{cosmo11} in the mass range from 10~keV$/c^2$ to 1~GeV$/c^2$. 

\section{\label{sec5} Conclusions and discussions.}
Light DM particles boosted by CRs are energetic and detectable in the sub-GeV mass region. Constraints on the $\chi$-nucleon scattering cross section are derived in this study based on the 205.4 kg day dataset using the CR boosting mechanism. More stringent limits are placed on the existing cosmological exclusion region for DM at a mass range from 10 keV$/c^2$ to 1 GeV$/c^2$.

During the CRDM calculation process, heavier CR nuclear components apart from the proton and helium are also taken into account, resulting in a doubling of the CRDM flux. Further, the unmodulated CR data with spatial distribution across the Galaxy generated by the $\tt GALPROP$ package are used. Previous CRDM studies used the modulated CR fluxes presented in Refs.~\cite{boschini_2017,Boschini_2018}, leading to conservative conclusions. Considering the inhomogeneity of the CR distribution, we use the $\tt GALPROP$ simulation package to calculate the value of $D_{eff}$ in this work and obtain a continuous function of the kinetic energy $T_{\chi}$ for different DM masses $m_{\chi}$, which is a more accurate approach compared with using a constant value~\cite{CRDM_PRD,CRDM_prl}.

During the Earth attenuation calculation, some studies~\cite{CRDM_prl,CRDM_TJ,CRDM_ZYF} adopted a ballistic-trajectory assumption to model the propagation of DM in Earth's crust, which neglects
the changing DM travel direction during the scattering process, leading to an extremely conservative conclusion. To improve the accuracy of the exclusion bound, we develop a Monte Carlo simulation package $\tt CJPL\_ESS$~\cite{cjpless}, in which the detailed topography of the Jinping Mountain is modeled, and the direction deviations of DM particles during their transportation are considered. The conventional Helm form factor is also adopted in the Earth attenuation simulation to evaluate the impact of the form factor, which depends on the energy scale of the momentum transfer during the scattering, whereas several works omitted it [$F(q)\sim 1$]~\cite{prospect,earthshielding,XQC1,CRDM_PRD}. The upper bound is elevated to $10^{-26}\ \rm cm^2$ due to these improvements. 

These results are obtained within the context of the $\chi$-N SI elastic scattering interaction. The constraints can be further improved after considering the CR-DM inelastic scattering process~\cite{Guo:inelastic,Bell:inelastic,Feng:inelastic}. The approach proposed in this work can be extended to new DM models, such as the gravitationally interacting DM~\cite{GIMP} and the highly interactie particle relics DM~\cite{HYPER_DM}. Advances in DM-electron scattering calculation in crystal targets~\cite{electron} allow new constraints on sub-GeV DM-electron scattering to be obtained in future work~\cite{CDEX_DM_e}. Furthermore, up-scattering of DM by the supernova shock waves~\cite{sndm1,sndm2} can also be probed.

\acknowledgments
We thank T. Bringmann and C. V. Cappiello for helpful discussions. This work was supported by the National Key Research and Development Program of China (Grant No. 2017YFA0402200) and the National Natural Science Foundation of China (Grants No. 12175112, No. 12005111, and No. 11725522). We acknowledge the Center of High Performance Computing, Tsinghua University, for providing the facility support.

\bibliography{CRDM}

\end{document}